\documentclass[number,sort&compress]{elsarticle}

\usepackage{amsmath}
\usepackage{hyperref}
\usepackage{textcomp}
\usepackage{lineno}
\usepackage{xcolor}
\usepackage{subcaption}

\begin{document}

\begin{frontmatter}

\title{Measurement of time resolution of the Mu2e LYSO calorimeter prototype}

\author[a]{N.~Atanov}
\author[a]{V.~Baranov}
\author[b]{F.~Colao}
\author[b]{M.~Cordelli}
\author[b]{G.~Corradi}
\author[b]{E.~Dan\'e}
\author[a]{Yu.I.~Davydov}
\author[c]{K.~Flood}
\author[b]{S.~Giovannella}
\author[a]{V.~Glagolev}
\author[b]{F.~Happacher}
\author[c]{D.G.~Hitlin}
\author[b,d]{M.~Martini}
\author[b]{S.~Miscetti\corref{cor2}}
\ead{stefano.miscetti@lnf.infn.it}
\author[c]{T.~Miyashita}
\author[e,f]{L.~Morescalchi}
\author[e,g]{G.~Pezzullo}
\author[b]{A.~Saputi}
\author[b]{I.~Sarra}
\author[b]{S.R.~Soleti\corref{cor1}}
\ead{soleti@lnf.infn.it}
\author[h,i]{G.~Tassielli}
\author[a]{V.~Tereshchenko}

\cortext[cor1]{Principal corresponding author}
\cortext[cor2]{Corresponding author}
\address[a]{Joint Institute for Nuclear Research, Dubna, Russia}
\address[b]{Laboratori Nazionali di Frascati, INFN, Frascati, Italy}
\address[c]{California Institute of Technology, Pasadena, United States}
\address[d]{Universit\`a ``Guglielmo Marconi'', Roma, Italy}
\address[e]{INFN, Pisa, Italy}
\address[f]{Universit\`a di Siena, Siena, Italy}
\address[g]{Universit\`a di Pisa, Pisa, Italy}
\address[h]{INFN, Lecce, Italy}
\address[i]{Universit\`a del Salento, Lecce, Italy}

\begin{abstract}
In this paper we present the time resolution measurements of the Lutetium-Yttrium Oxyorthosilicate (LYSO) calorimeter prototype for the Mu2e experiment. The measurements have been performed using the $e^-$~beam of the Beam Test Facility (BTF) in Frascati, Italy in the energy range from 100 to 400 MeV.
The calorimeter prototype consisted of twenty five 30x30x130~mm$^3$, LYSO crystals read out by 10x10 mm$^2$ Hamamatsu Avalanche Photodiodes (APDs).
The energy dependence of the measured time resolution can be parametrized as $\sigma_{t}(E)=a/\sqrt{E/\mathrm{GeV}} \oplus b$, with the stochastic and constant terms $a=(51\pm1)$~ps and $b=(10\pm4)$~ps, respectively. This corresponds to the time resolution of ($162\pm4$)~ps at 100~MeV. 
\end{abstract}


\end{frontmatter}
\pagebreak{}
\section{Introduction}
The Mu2e experiment at Fermilab \cite{Mu2e_tdr} aims to search for Charged Lepton Flavor Violation (CLFV) in the neutrinoless, coherent conversion of a negative muon into an electron in the Coulomb field of an $^{27}$Al nucleus. The $\mu\rightarrow e$ conversion results in monoenergetic electrons with an energy equal to the muon rest mass minus the corrections for the nuclear recoil and the binding energy of the muon. For $^{27}$Al, the energy of the monoenergetic electron is $E_e$ = $104.97$~MeV~\cite{spectrum}.

The experiment is designed to reach the single event sensitivity (SES) of $2.4\times10^{-17}$ in three years of running \cite{Mu2e_tdr}. This value represents an improvement of four orders of magnitude over the current best experimental limit $R_{\mu e}(\mathrm{Au})$~$<$~7$\times$10$^{-13}$ @ 90\% C.L. set by the SINDRUM II experiment \cite{Sindrum}.

The Standard Model predicted rate for this process is $\mathcal{O}(10^{-52})$ \cite{clfv}, therefore any signal observed by Mu2e would be a compelling evidence of new physics.

\section{The Mu2e electromagnetic calorimeter}

The Mu2e detector is designed to identify $\mu\rightarrow e$ conversion electrons and reduce the background to a negligible level. The detector is located inside a large superconducting solenoid with the magnetic field $B$=1 T and surrounded by the cosmic ray veto counters. 

A low mass straw tracker provides an accurate track momentum measurement necessary to separate the signal from the background. The calorimeter is located behind the tracker and  complements it by providing: 
(i)  powerful $\mu$/e particle identification (PID), 
(ii) seeds for the pattern recognition in the tracker 
and (iii) an independent software trigger system.

Efficient PID requires the calorimeter to have the timing resolution  better than 500 ps, the energy resolution of $\mathcal{O}$(5\%) is needed to provide an efficient trigger. The calorimeter should be able to operate in an environment where a radiation dose up to 120 Gy/year is delivered by protons, neutrons, and photons. It must also function  in a 1 T axial magnetic  field and a $10^{-4}$ Torr vacuum. Before the sudden increase of the lutetium price, the Mu2e calorimeter design included two disks of LYSO crystals read out by two large area avalanche photodiodes (APDs) per crystal \cite{cdr}. The choice of LYSO as a scintillator provided high light yield, fast response, and radiation hardness\cite{zhou, zhou2, dissertori}. 

In this paper, we report the results of tests performed with a LYSO-based calorimeter prototype, which include measurements of the timing resolution and evaluation of the front-end electronics (FEE) and readout system.

\section{The LYSO crystal calorimeter prototype}\label{aba:sec3}

\begin{figure}
\centering
\includegraphics[width=0.45\linewidth]{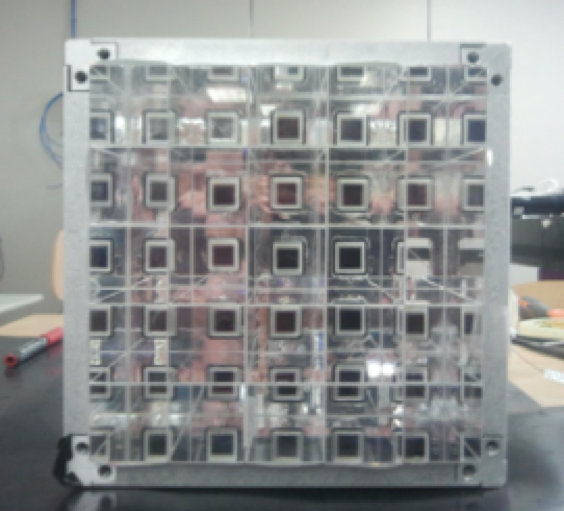}
\includegraphics[width=0.4655\linewidth]{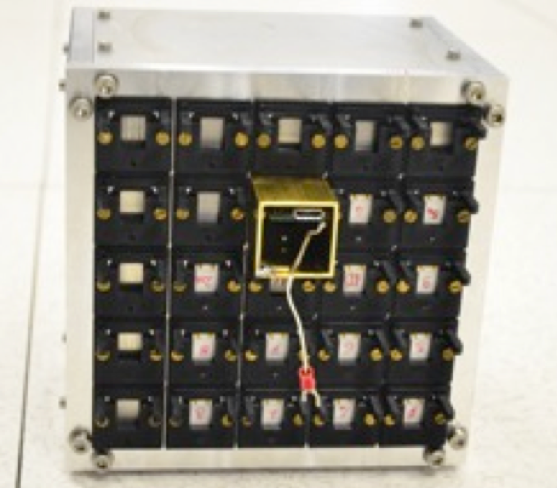}

\caption{On the left (right), is the front (back) view of the 5x5 LYSO calorimeter prototype. The APDs that are attached to the back of each crystal are visible in both views. The Fig. on the right shows a brass Faraday cup that is placed around placed around each Amp-HV board.}
\label{fig:matrix}
\end{figure}

The calorimeter prototype consisted of 25 LYSO crystals (30x30x130 mm$^3$) from the Shanghai Institute of Ceramics, Chinese Academy of Sciences (SICCAS) \cite{siccas}, arranged in a 5x5 matrix (Fig.~\ref{fig:matrix}). Each crystal was identified by two indices ($i$,$j$), corresponding to its row and column positions, with the crystal (0,0) being the bottom-left crystal in the matrix viewed from the back. The dimensions corresponded to $\sim$ 11.2 radiation lengths (X$_0$) in depth and a transverse size of $\sim$ 3.6 Moli\`ere radius (R$_M$). 
The crystals were individually characterized with a $^{22}$Na source and a spectrophotometer, 15 crystals at INFN Laboratori Nazionali di Frascati (LNF) and 10 at the California Institute of Technology (Caltech).  All crystals demonstrated high light yield and transmittance. The measured longitudinal response uniformity was below 5\%. Each crystal was wrapped in a 60~\textmu m thick layer of a super-reflective 3M Enhanced Specular Reflector (ESR) film \cite{3m} and read out by a Hamamatsu S8664-1010 APD \cite{hamamatsu}. The APDs were optically connected to the crystals with Saint-Gobain BC-630 optical grease \cite{saintgobain}. 

\begin{figure}
\centering
\includegraphics[width=0.4745\linewidth]{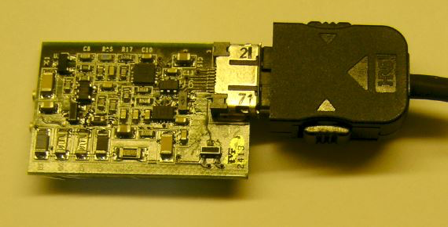}
\includegraphics[width=0.45\linewidth]{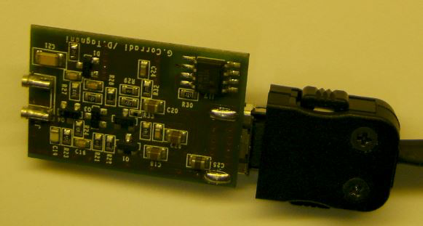}

\caption{A picture of an Amp-HV prototype. Left: amplification. Right: HV side.}
\label{fig:FEE}
\end{figure}

The Front-End Electronics (FEE) consisted of a multi-layer, double-sided, discrete board (Amp-HV) directly connected to the photosensor pins (see Fig.~\ref{fig:FEE}). 
The board provided both the amplification stage and the local regulation of the photosensor bias voltage, thus reducing the noise loop-area.
The amplification layer was a double stage transimpedance preamplifier with a total gain of 15~k$\Omega$, which maintained an equivalent noise charge (ENC) level of about 10$^3$ electrons with no input capacitance source. The linear regulation layer allowed precise voltage regulation and long-term stability of better than 100 ppm. Each group of 16 Amp-HV chips was controlled by an external ARM controller distributing the voltages. Two ARM controllers were used for the prototype.
The high voltage (530~V) was produced by a primary generator that used the low-noise switching technology and resided on the ARM controller board. The output voltage was regulated by a DAC and read out with a 16-bit ADC.

APDs were calibrated using a green (530~nm) 0.6~\textmu J/pulse laser. The laser light was distributed through 250 \textmu m core diameter fused silica optical fibers. The fibers were inserted into a custom connector, polished and positioned directly in the APD holders. The 10~ns laser pulse was synchronized with an external trigger.

\section{Experimental setup at BTF}\label{sec:setup}

\begin{figure}
\begin{center}
\includegraphics[scale=0.24]{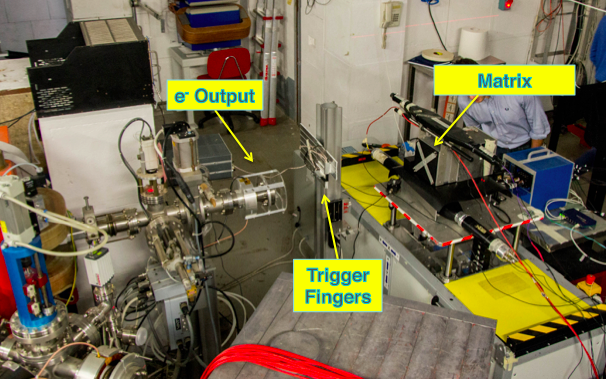}
\includegraphics[scale=0.29]{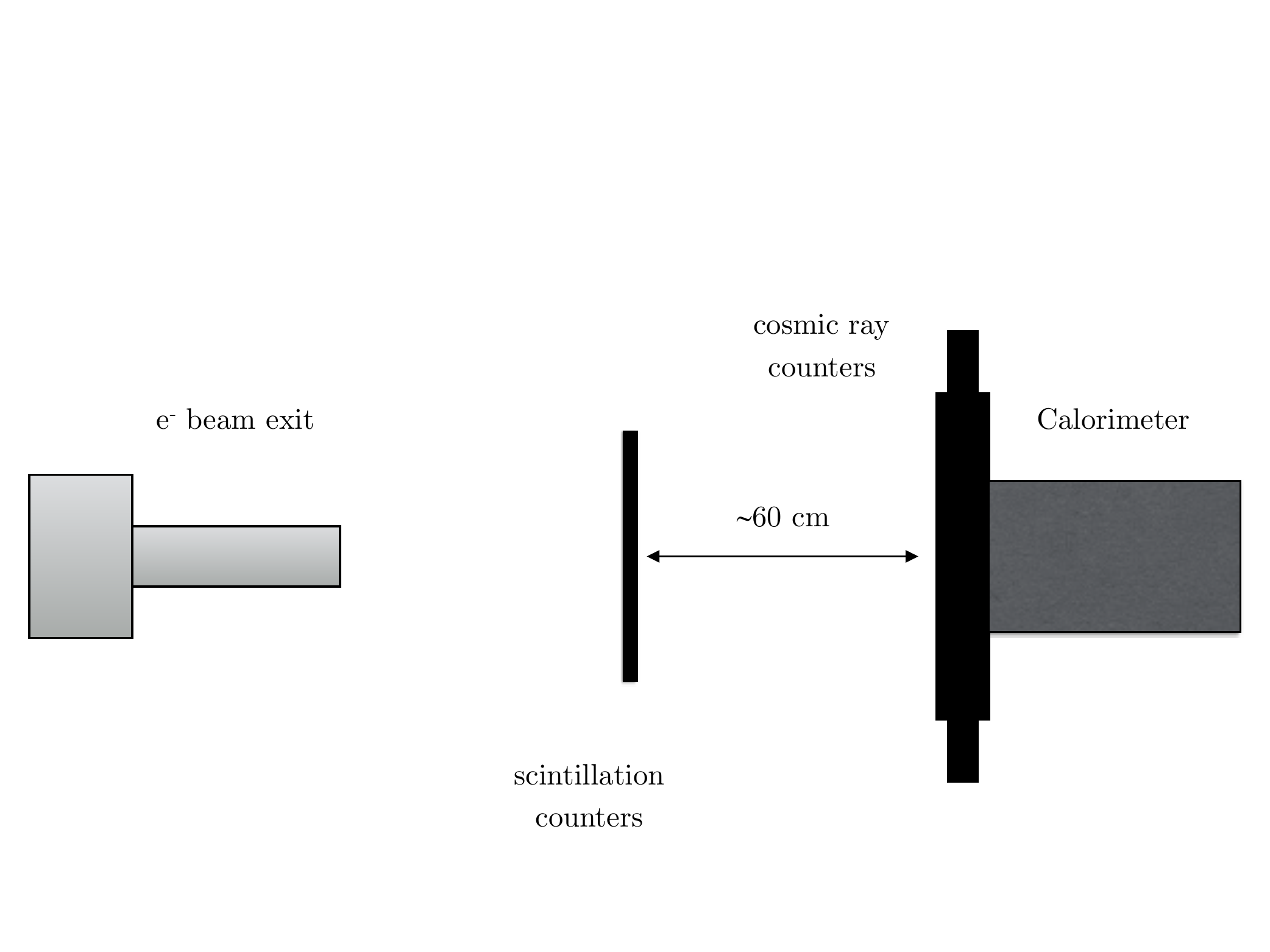}

\caption{Left: picture of the BTF hall with the experimental setup used for the calorimeter prototype time resolution measurement. Right: schematic diagram of the experimental setup seen from the top.}
\label{fig:btfhall}
\end{center}
\end{figure}

The calorimeter prototype was tested at the Beam Test Facility~\cite{BTF} (BTF) of the INFN LNF in December 2014 and again in April 2015. The BTF is a part of the DA$\mathrm{\Phi}$NE (Double Annular Factory for Nice Experiments) accelerator complex equipped for testing particle detectors. The complex includes a Linac which sends the beam pulses to the BTF area at a 50 Hz frequency. 
Each Linac pulse has a $\sim$10 ns duration and is divided into 180-200~ps long bunches. The trigger signal provided by the Linac has a time resolution on the $\mathcal{O}$(10 ns) time resolution, making it necessary to use a different source of the trigger timing. Two 6x10x50 mm$^{3}$ finger-shaped beam scintillation counters located upstream of the calorimeter prototype have been used for this purpose. Due to interference with other detectors and logistic in the area, we were unable to place the counters closer than 60~cm to the calorimeter surface. The scintillation counters were read out by 3x3 mm$^{2}$ SensL \cite{sensl} silicon photomultipliers (SiPMs) (Fig.~\ref{fig:btfhall}).

In order to select cosmic rays, two plastic scintillation counters, 50x50x200 mm$^3$ in size, were positioned above and below the calorimeter prototype. Each of those cosmic counters has been read out by two photomultipliers (PMTs). 

Initial channel-to-channel calibration was performed with cosmic ray minimum ionizing particles (MIPs) and later updated using more accurate beam calibration.

The data taking configuration used an OR of three different triggers: (1) a beam trigger (BT) formed by the AND of signals from the finger scintillation counters. 
The rate of this trigger varied from run to run, from a few Hz up to 20 Hz; (2) a cosmic ray trigger (CRT). The rate of this trigger was at a level of 2 Hz; (3) a laser trigger (LT) generated by a timer at a typical rate of 0.1 Hz.
The CRT and LT triggers were also used to monitor the calorimeter response during the data taking.

Data were acquired by a VME-based DAQ system from CAEN \cite{caen} and read out by 4 CAEN V1720 waveform digitizers at a sampling rate of 250 Msps with 12-bit resolution over the 0-2 V dynamic range.

For each calorimeter channel and each trigger, the signal charge was determined by numerical integration of the signal waveform in a gate of [-50, 450]~ns around the time sample corresponding to the pulse maximum. The charge baseline was estimated by integrating the waveform in a gate [-750, -250]~ns before the pulse maximum. The baseline was then subtracted from the signal on event by event basis.

\section{Equalization and calibration}

In order to measure the time resolution of the calorimeter prototype in the energy range from 100 to 400 MeV, two different APD gain settings were used. A gain of $G=75$ was used for runs at 100, 150 and 200 MeV. For signals to stay within the dynamic range of the digitizer a gain of $G=25$ was used at higher beam energies. The laser trigger was used to adjust gains in individual channels. Waveforms from each channel were monitored with a software scope and the pulse heights were equalized by adjusting the individual APD HV settings.



The crystal-to-crystal response was determined by directing 450 MeV electrons onto the center of each crystal.  The global charge-to-energy conversion scale was set by comparing the total energy reconstructed in the calorimeter, $E_{rec}$, to the expected energy deposition estimated by the Monte Carlo (MC) simulation based on \texttt{GEANT4} \cite{geant}.

\section{Event selection}
\begin{figure}[htbp]
\begin{center}
\includegraphics[scale=0.28]{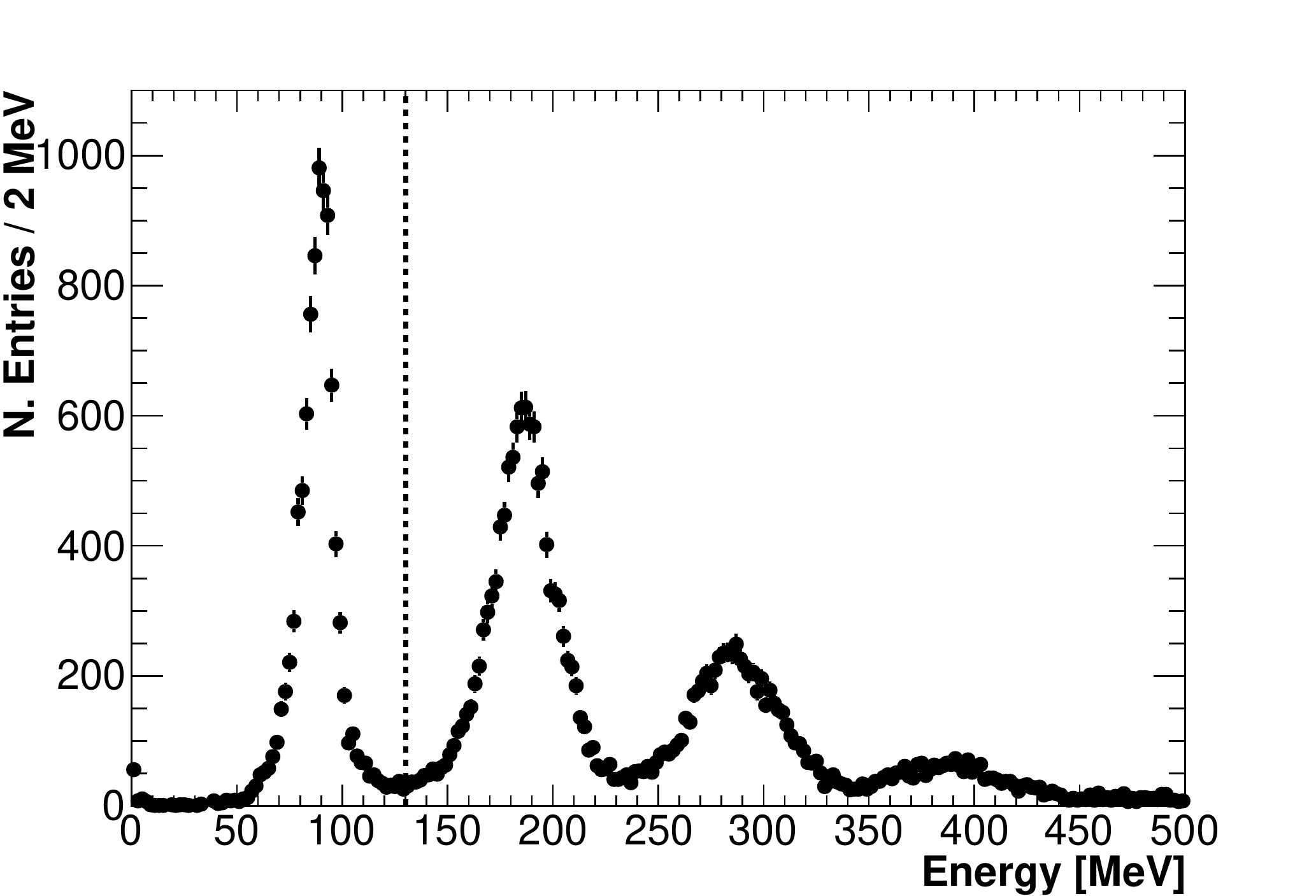}
\includegraphics[scale=0.28]{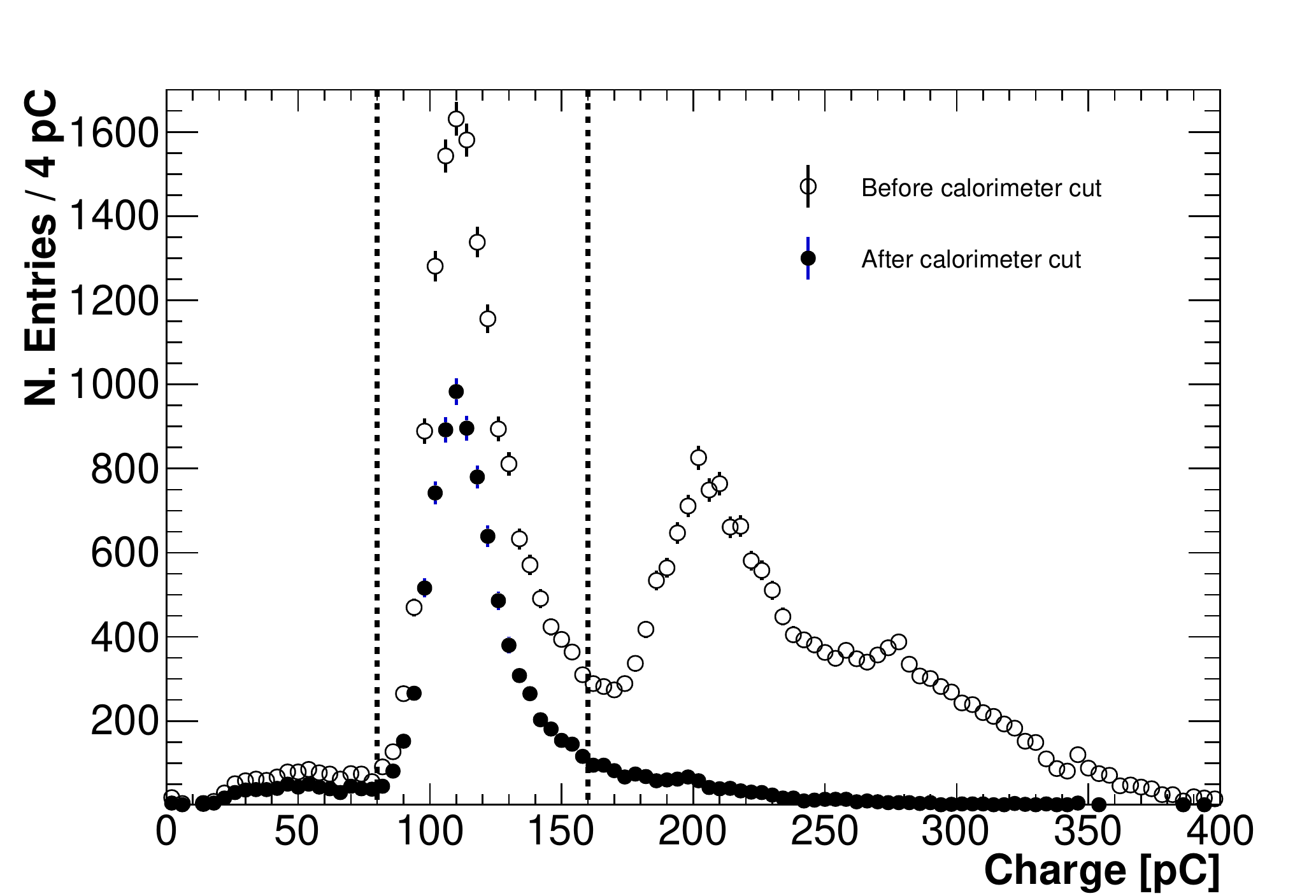}

\caption{Left: reconstructed energy in the calorimeter prototype for an $e^-$ beam of 100 MeV. Right: charge response for one of the scintillator counters for an $e^-$ beam of 100 MeV before and after the cut on the energy deposited in the calorimeter. The single-electron selection cut has been set to $E_{rec}<1.3\cdot E_{beam}$ for the reconstructed energy in the calorimeter prototype and to $80 \thinspace\mathrm{pC} < Q_{dep} < 160 \thinspace\mathrm{pC}$ for the charge deposition in the counter, as shown by the dashed lines.}
\label{fig:charge_cuts}
\end{center}
\end{figure}

The BTF beam intensity was tuned to provide the mean number of electrons per bunch $\lambda\simeq$ 0.8 at 100 MeV. At this intensity a non-negligible fraction of bunches contains two or more electrons. In order to select only single-electron events, a cut on the total reconstructed energy in the calorimeter prototype, $E_{rec}$,  and the signal charge in each of the scintillation counters, $Q_{dep}$, has been applied.
The cut on the energy deposited in the calorimeter prototype has been set to $E_{rec}<1.3\cdot E_{beam}$, as shown in Fig.~\ref{fig:charge_cuts}(left). An example of a $Q_{dep}$ cut is shown in Fig.~\ref{fig:charge_cuts}(right).

\subsection{Centroid cut}
\begin{figure}[htbp]
\begin{center}
\includegraphics[scale=0.5]{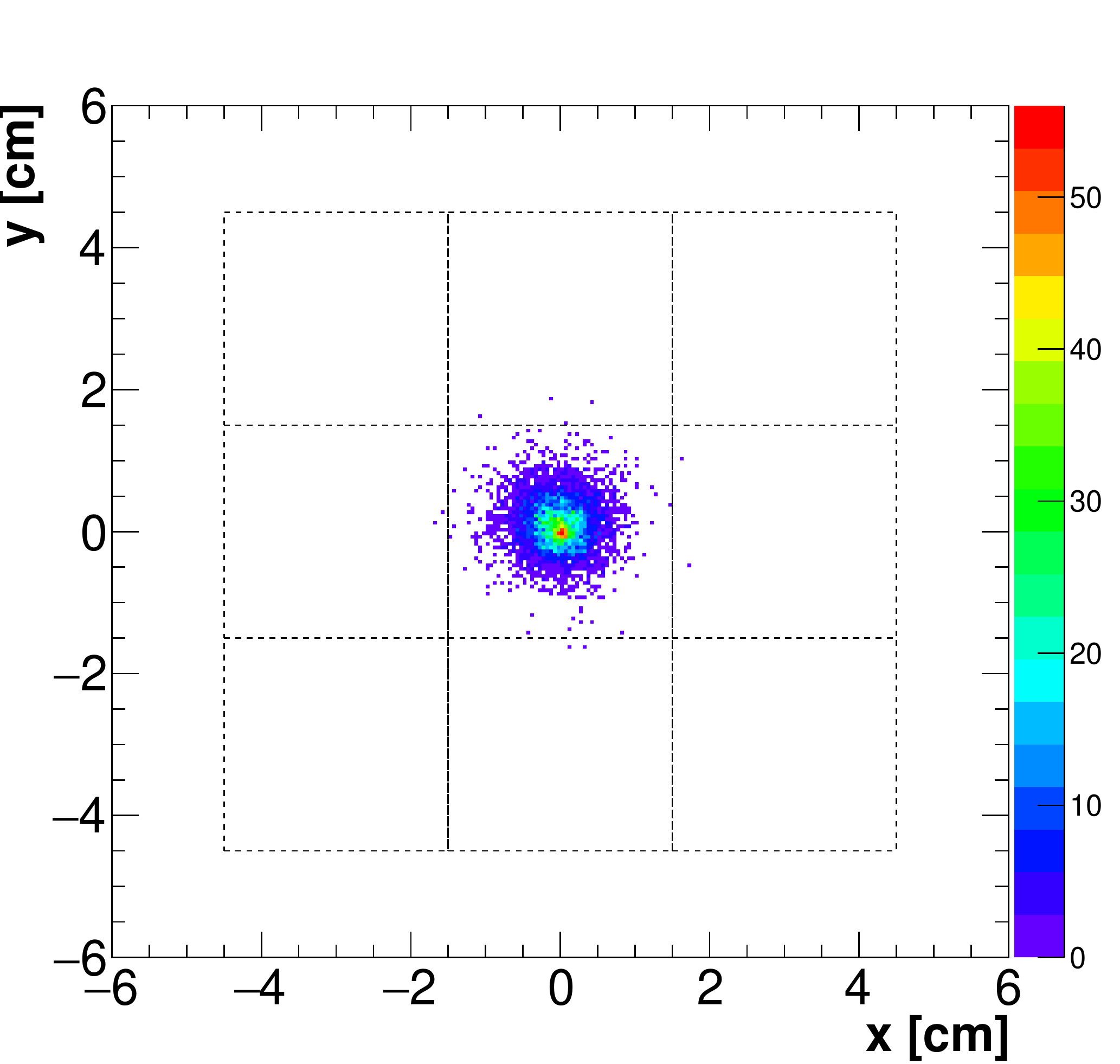}

\caption{Spatial distribution of hits on the front face of the calorimeter prototype at $E_{beam}=100$~MeV. The dashed lines show 9 of 25 crystals. }
\label{fig:centroid}
\end{center}
\end{figure}

Because of $\sim$60~cm distance between the calorimeter prototype and the scintillation counters, the effect of the electron multiple Coulomb scattering downstream of the counters was not negligible, especially at low energies. In Fig.~\ref{fig:centroid} the correlation between the $x$ (horizontal) and $y$ (vertical) coordinates reconstructed  in the calorimeter  for $E_{beam}=100$~MeV is shown.  The $x$ ($y$) coordinate is given by the logarithmic energy-weighted average of the crystal positions
\begin{equation}
x = \frac{\sum\limits^{25}_{i=1}x_{i}\left[w_{0}-\mathrm{log}(E_{i}/E_{tot})\right]}{\sum\limits^{25}_{i=1}\left[w_{0}-\mathrm{log}(E_{i}/E_{tot})\right]},
\end{equation}
where
$x_i$ can be 0 cm, $\pm$3 cm, $\pm$6 cm (centers of the crystals with respect to the beam line), $E_i$ is the energy deposition in the $i$-th crystal, $E_{tot}$ is the total energy deposited in the calorimeter and $w_0$ is a custom parameter set to 9.  
%
To suppress effects of multiple scattering we require
\begin{equation}
r  = \sqrt{x^2+y^2} < 0.5 \thinspace\mathrm{cm}.
\end{equation}
As the multiple scattering scales as $1/p$, at higher beam energies its effect becomes smaller. 
\section{Determination of the time resolution}

\begin{figure}[htbp]
\begin{center}
\includegraphics[scale=0.28]{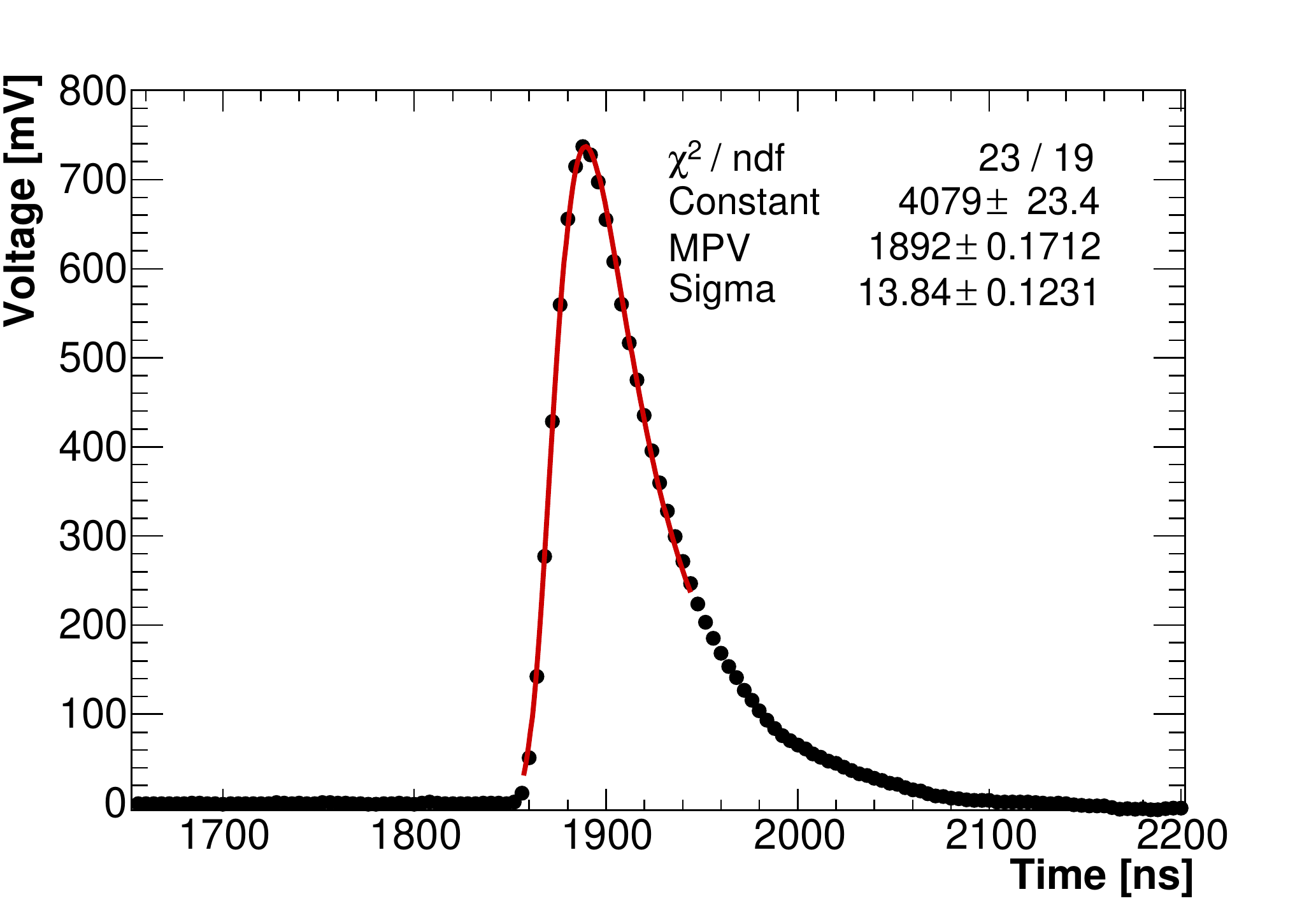}
\includegraphics[scale=0.28]{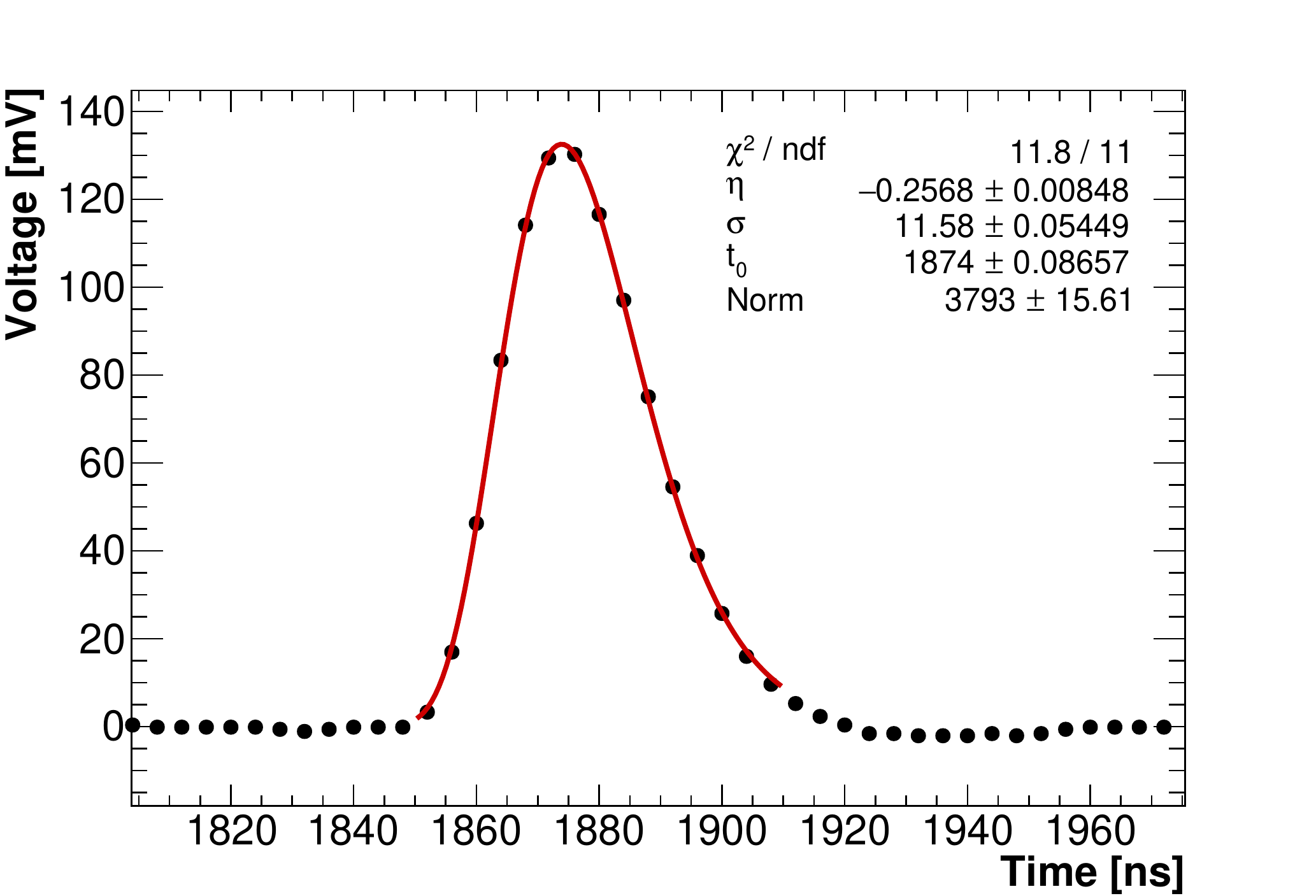}

\caption{Examples of the calorimeter signals. Left: waveform for a $e^-$ beam event fitted with the Landau distribution. Right: waveform for a laser event fitted with the log-normal distribution.}
\label{fig:waveforms}
\end{center}
\end{figure}

To extract the calorimeter timing information, the waveform signals associated with $e^-$ events have been fitted with a Landau function in the range $[t_{max}-30\thinspace\mathrm{ns},\thinspace t_{max}+70\thinspace\mathrm{ns}]$, where $t_{max}$ is the time of the waveform peak, as measured by the digitizer (Fig.~\ref{fig:waveforms}). Signals from the trigger counters have been fitted with a log-normal distribution \cite{logN} in the range $[t_{max}-15\thinspace\mathrm{ns},\thinspace t_{max}+10\thinspace\mathrm{ns}]$.



The time corresponding to the maximum of the fitted function has been used as the signal time. Signals above 10 mV were used in the data analysis.  



\subsection{Calorimeter time resolution with an external start}\label{sec:counters}
The time resolution was determined from the width of $\Delta t = t_{signal} - t_{start}$ distributions, using either the central crystal alone or the entire calorimeter prototype. In the latter case, $t_{signal}$ is an energy-weighted sum of times reconstructed in different channels:
\begin{equation}
t_{signal} = \frac{\sum\limits^{25}_{i=1}t_{i}E_{i}}{\sum\limits^{25}_{i=1}E_{i}},
\label{eq:tmatrix}
\end{equation}
where
$t_i$ is the peak time of the $i$-th crystal, $E_i$ is the energy deposition in the $i$-th crystal, and $t_{start} = (t_{f_{1}}+t_{f_{2}})/2$ is the average of the beam counter times.

\begin{figure}[htbp]
\begin{center}
\includegraphics[scale=0.6]{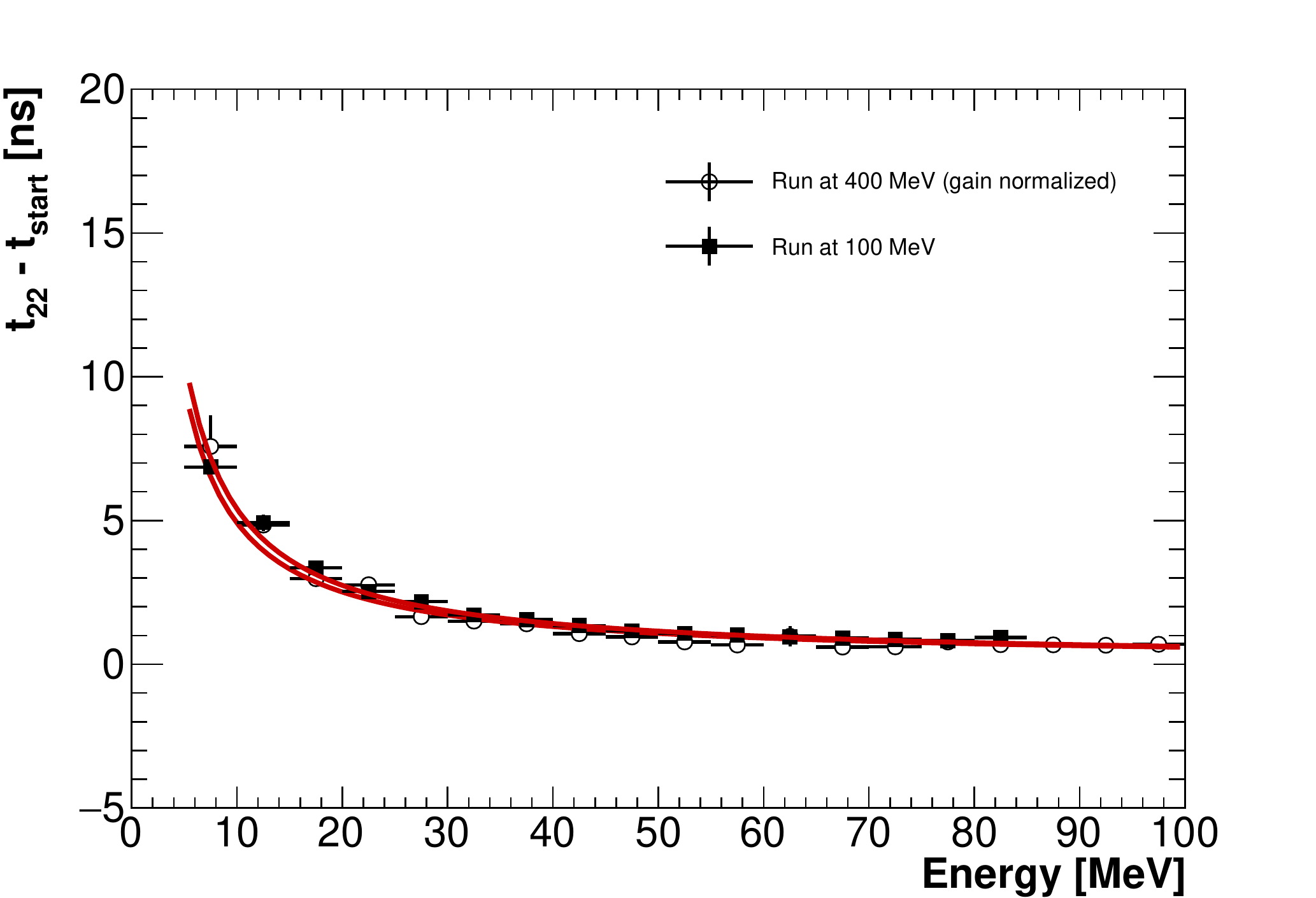}

\caption{Central crystal timing ($t_{22}$) as a function of the deposited energy 
  in the central crystal for $E_{beam}$ = 100 MeV and $E_{beam}$ =
  400 MeV. Energy deposition for $E_{beam}$ =
  400 MeV has been corrected by a factor of 3 due to the different gains ($75/25$).}
\label{fig:slewing}
\end{center}
\end{figure}

After correcting for delays due to the cable length differences between channels ($T^0$s), a residual time-walk effect, a dependence of $t_{i}$ on the deposited energy, remains. 
In Fig.~\ref{fig:slewing}, examples of this dependence for the central crystal for $E_{beam}$ = 100 MeV and $E_{beam}$ = 400 MeV runs are shown. Due to the different gains (25 vs 75), the energy depositions at $E_{beam}$ = 400 MeV run have been scaled down by a factor of 3. We parameterize the dependence with a $a+b/E+c/\sqrt{E}$ function and correct the individual reconstructed times in Eq. \eqref{eq:tmatrix} as follows:
\begin{equation}
t_{i}^{*}= t_i - a_i - \frac{b_i}{E_i} - \frac{c_i}{\sqrt{E}}.
\end{equation}

\begin{figure}[htbp]
\begin{center}
\begin{subfigure}{.45\textwidth}
\includegraphics[scale=0.28]{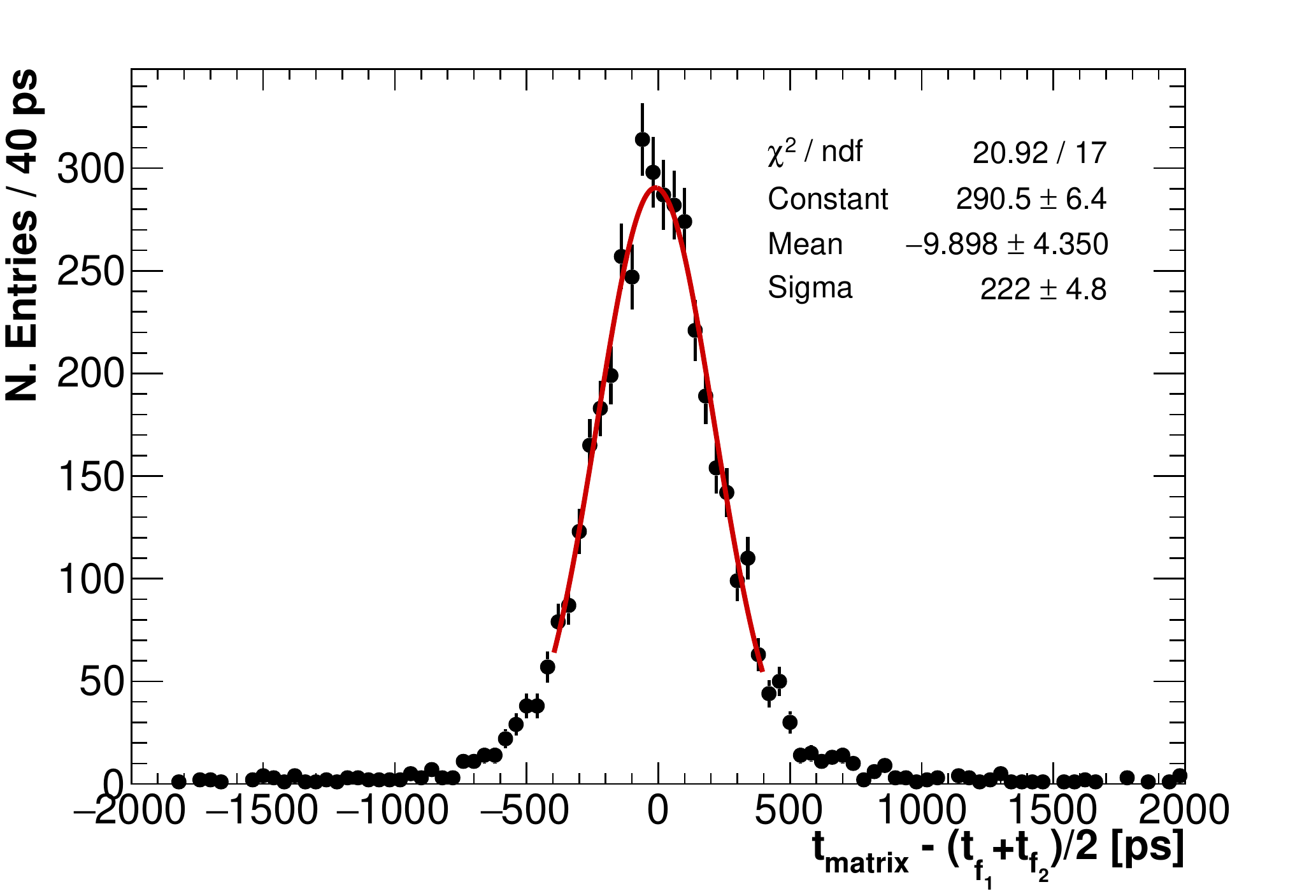}
\caption{100 MeV}
\end{subfigure}
\begin{subfigure}{.45\textwidth}
\includegraphics[scale=0.28]{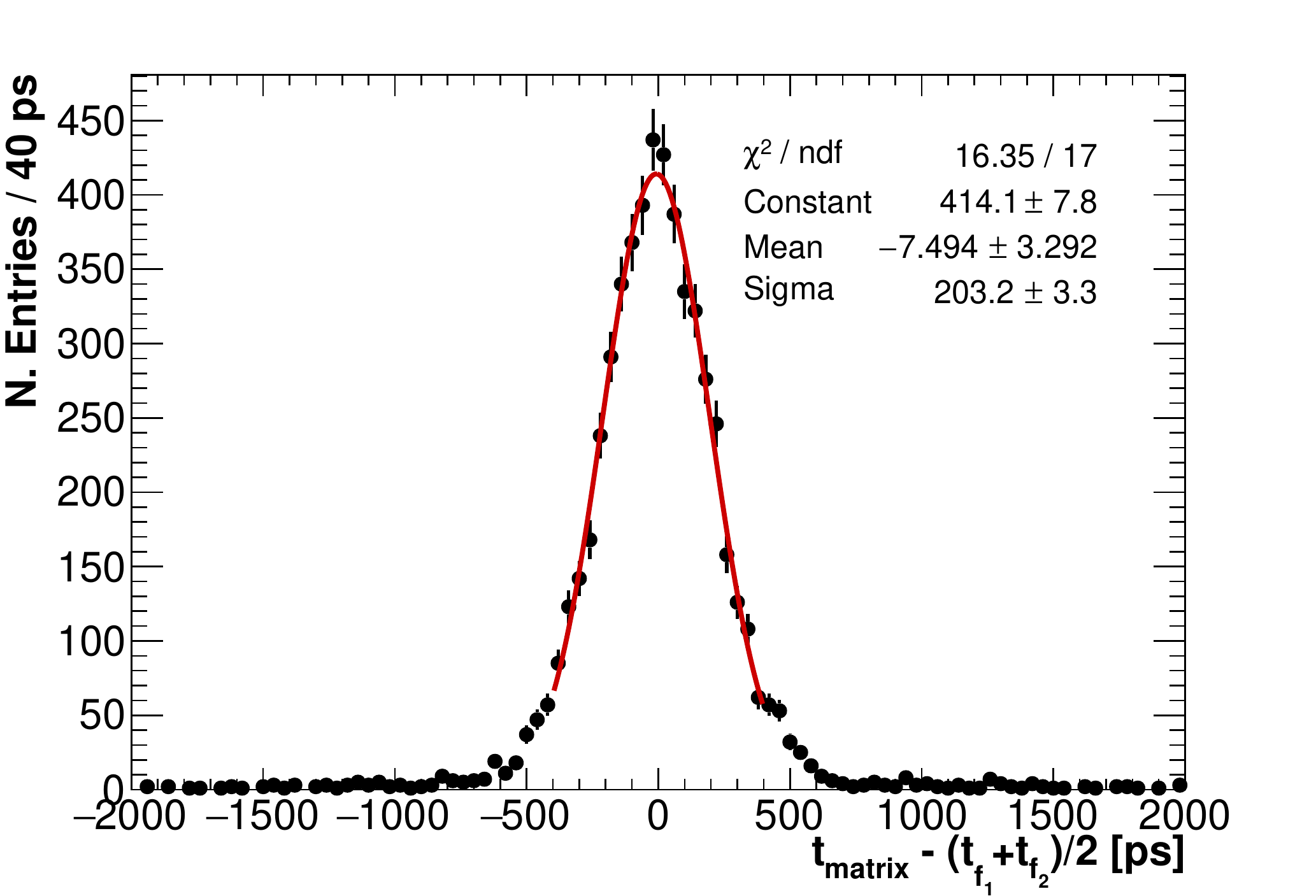}
\caption{200 MeV}
\end{subfigure}
\begin{subfigure}{.45\textwidth}
\includegraphics[scale=0.28]{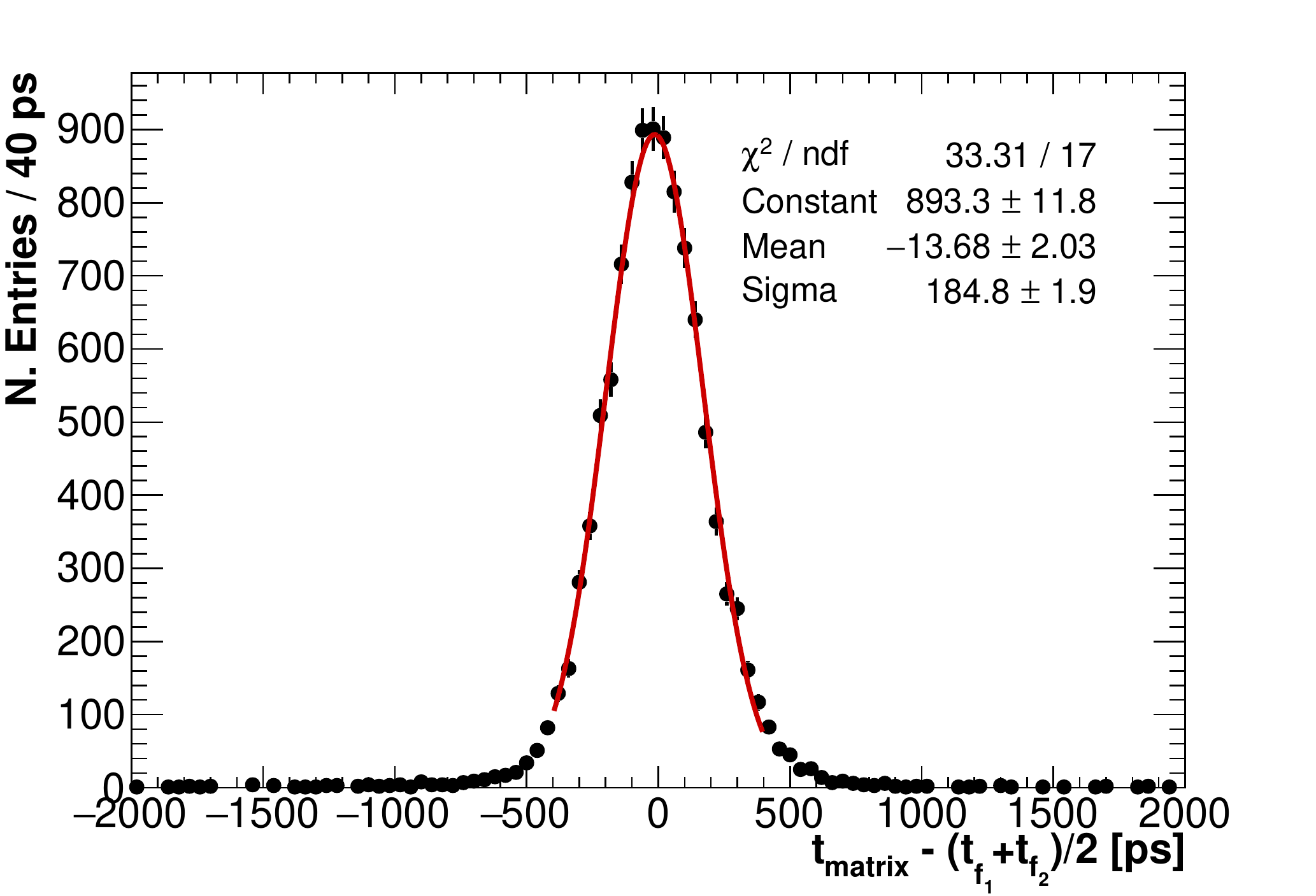}
\caption{300 MeV}
\end{subfigure}
\begin{subfigure}{.45\textwidth}
\includegraphics[scale=0.28]{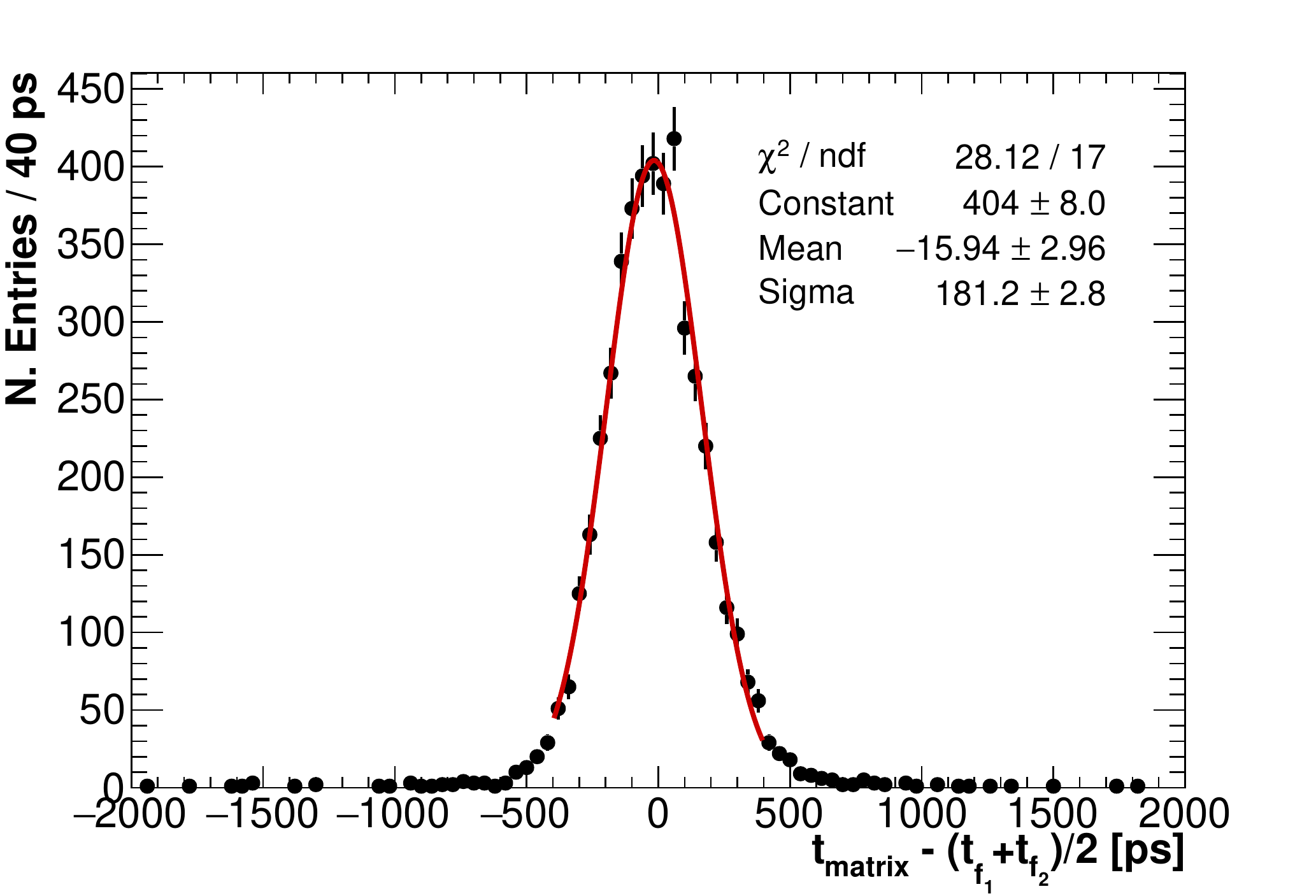}
\caption{400 MeV}
\end{subfigure}

\caption{Corrected distributions of $t_{matrix}$ for 100, 200, 300, and 400 MeV. The long tail present on both the high and low sides is related to a small contamination of two-particle events that are still present in the selected sample. However, the high-side tail in the time distribution at 200 MeV (top-right) is related to the pulse height  saturation observed at that energy.}
\label{fig:timing}
\end{center}
\end{figure}

Fig.~\ref{fig:timing} shows distributions of the corrected $t_{signal}$ at 100, 200, 300, and 400 MeV with the gaussian fits superimposed. Gaussians describe well the central part of the distributions,  a small contribution of 2-electron events leads to non-gaussian tails. By observing the relative shifts of the means of the distributions with respect to zero, the precision of the calibration procedure was estimated to be better than 20 ps.

\begin{figure}[htbp]
\begin{center}
\includegraphics[scale=0.6]{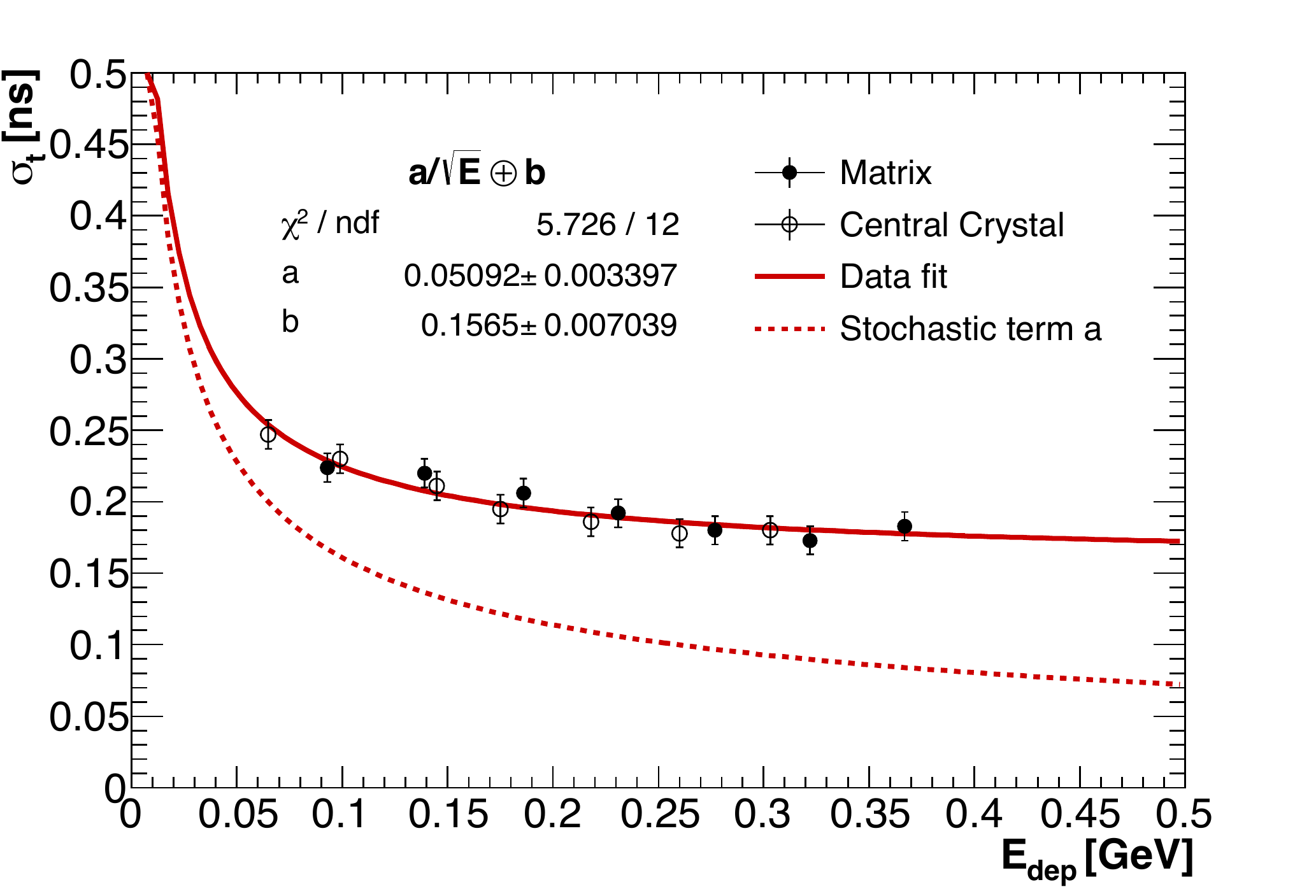}

\caption{Time resolution, as defined by the central crystal and the energy-weighted sum 
  over all crystals, as a function of the deposited energy $E_{dep}$. The
  dashed line represents the stochastic term only.}
\label{fig:jitter}
\end{center}
\end{figure}

The dependence of the time resolution on $E_{dep}$ is shown in Fig.~\ref{fig:jitter} and is well described by the function
\begin{equation}
\label{eq:res}
\sigma_{t}(E_{dep}) = \frac{a}{\sqrt{E_{dep}/\mathrm{GeV}}}\oplus b,
\end{equation}

\noindent
where $a$ is the stochastic term and $b$ is the constant term, which is mainly due to the trigger jitter. From the fit, $a= (51\pm3)$ ps and $b= (157 \pm 7)$ ps.

\begin{figure}[htbp]
\begin{center}
\includegraphics[scale=0.6]{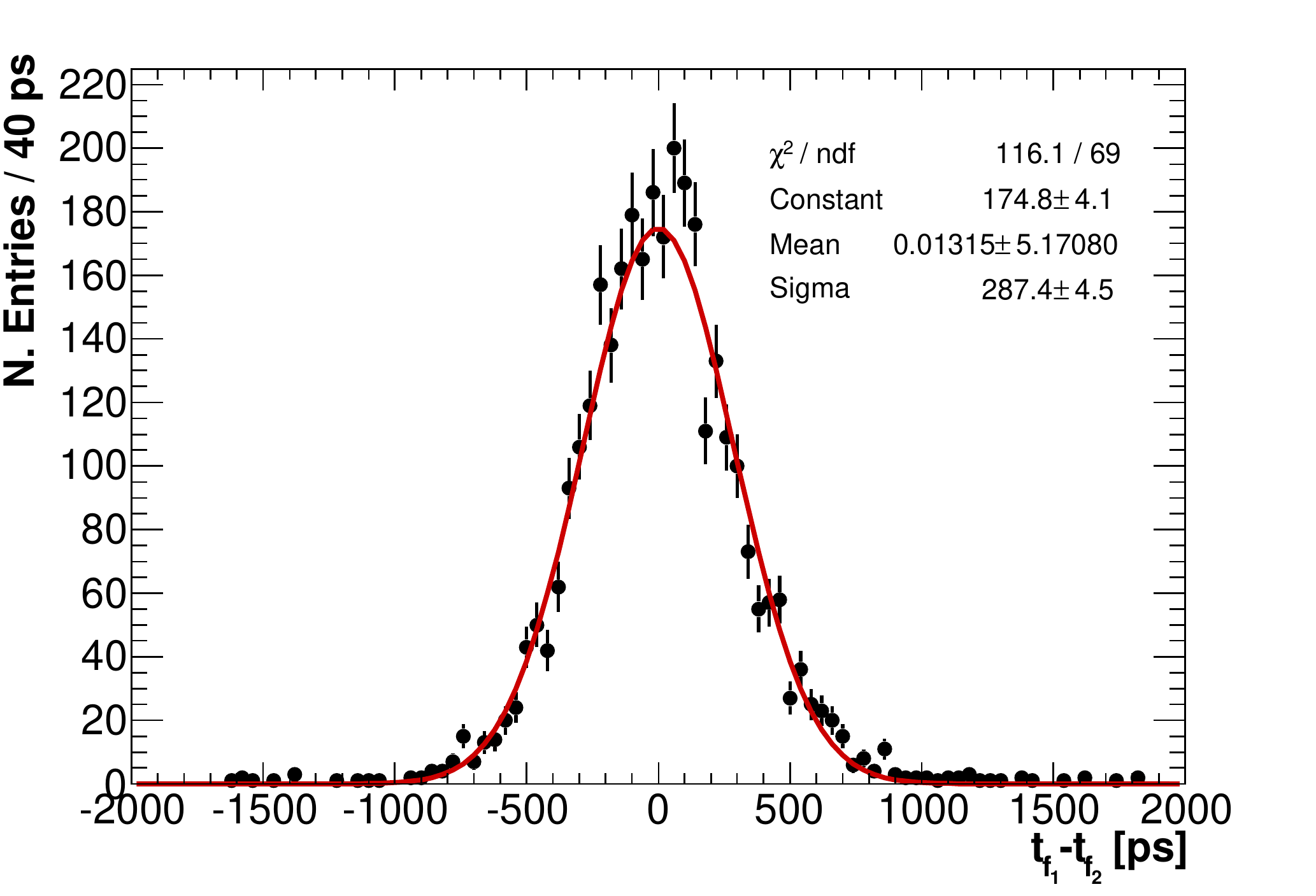}

\caption{$t_{f_{1}}-t_{f_{2}}$ distribution for a 100 MeV $e^-$ beam. The half width of the gaussian gives an estimate of the trigger jitter.}
\label{fig:delta_t_gauss_fit}
\end{center}
\end{figure}

The trigger jitter can be estimated directly by fitting the distribution of $t_{f_1} - t_{f_2}$ with a gaussian, as shown in Figure~\ref{fig:delta_t_gauss_fit}. The width of the distribution returned by the fit is $\sigma = (287 \pm 5)$ ps.  Assuming the resolution of both counters is the same
\begin{equation}
\label{eq:aaa}
\sigma\left(t_{start}\right) = \frac{1}{2}\sigma(t_{f_{1}}-t_{f_{2}}) = 144 \pm 3\thinspace\rm{ps},
\end{equation}
consistent with the constant term from the energy dependence fit \eqref{eq:res}.

\subsection{Calorimeter-based time resolution}
The calorimeter time resolution can also be determined by measuring the time difference between the signals in the two neighboring crystals. This technique does not require an external time reference and is widely used in HEP~\cite{cms}. This method has been applied to the data collected at 100 and 200 MeV with the beam offset horizontally by 0.6, 1.0 and 1.5 cm with respect to the prototype center.
Events with the reconstructed energies in the neighboring crystals $E_{12}$ and $E_{22}$
satisfying the requirement:
\begin{equation}
  0.8 <  E_{12}/E_{22} < 1.2, 
  \end{equation}
have been selected for analysis.
Time-walk corrections in individual channels have been determined as described in section~\ref{sec:counters}.

\begin{figure}[htbp]
\begin{center}
\includegraphics[scale=0.6]{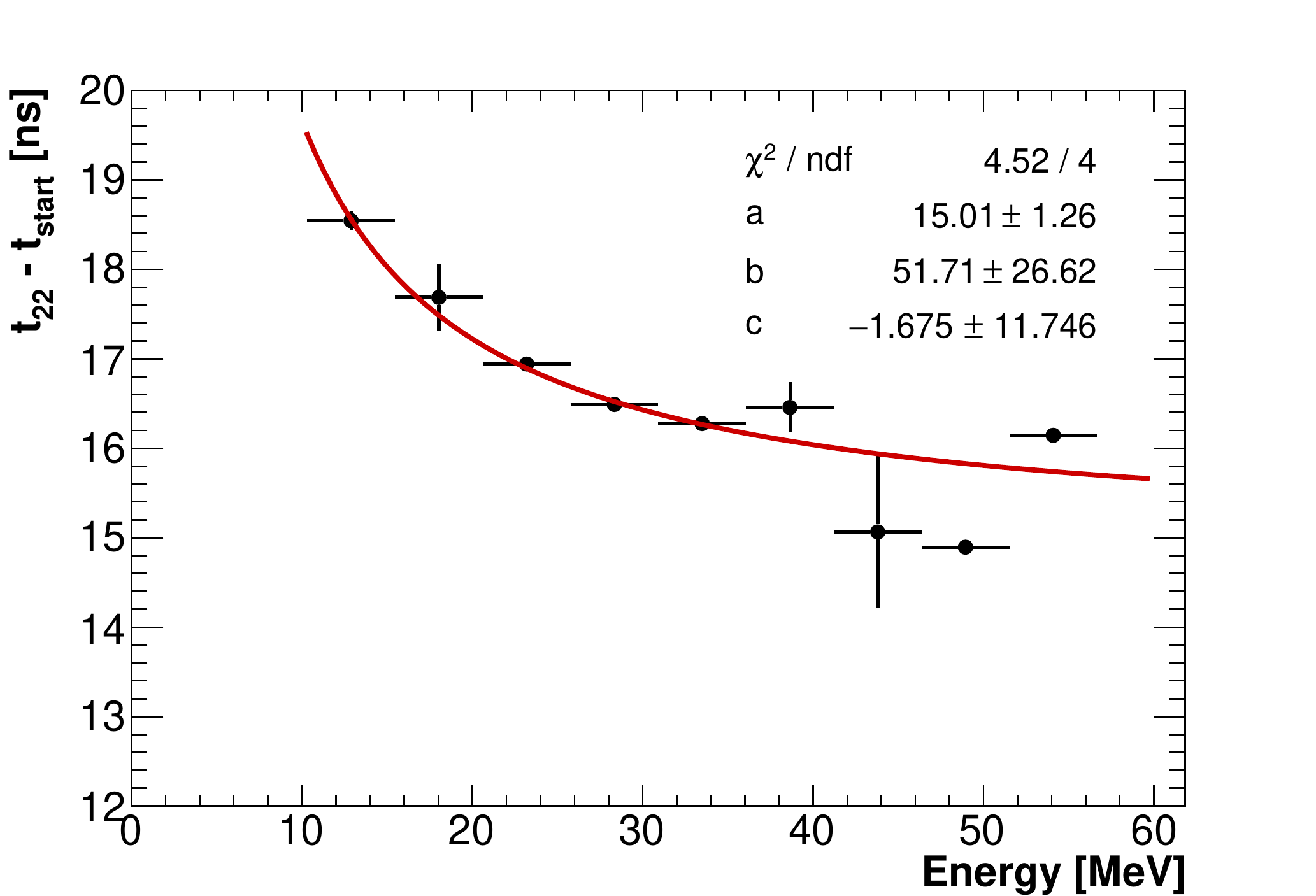}

\caption{Time-walk corrections for the run at 100 
  MeV. $t_{22}$ is the crystal timing measurements
  for the central crystal before the correction. The fit function is described in section~\ref{sec:counters}.}
\label{fig:slewinsitu}
\end{center}
\end{figure}
\begin{figure}[htbp]
\begin{center}
\includegraphics[scale=0.28]{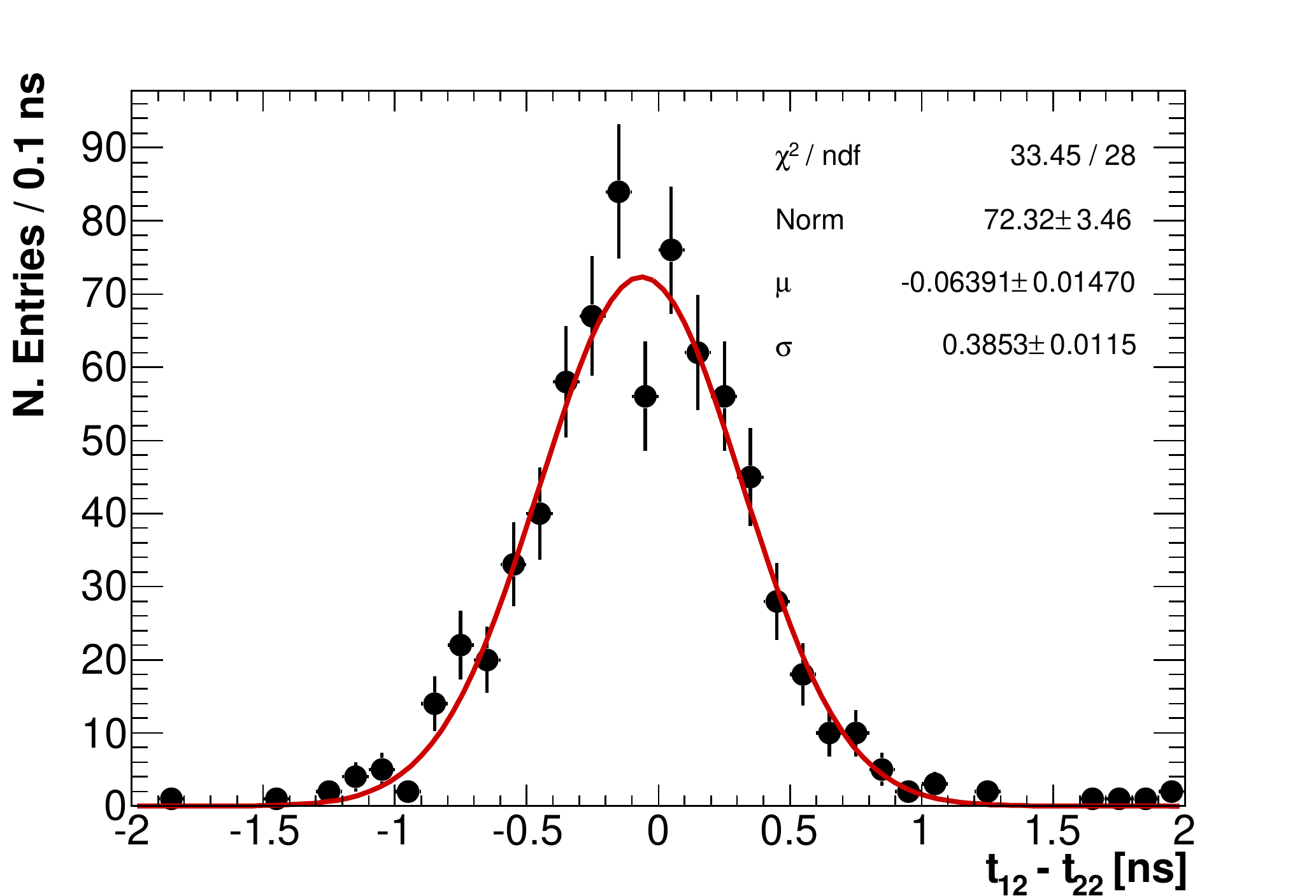}
\includegraphics[scale=0.28]{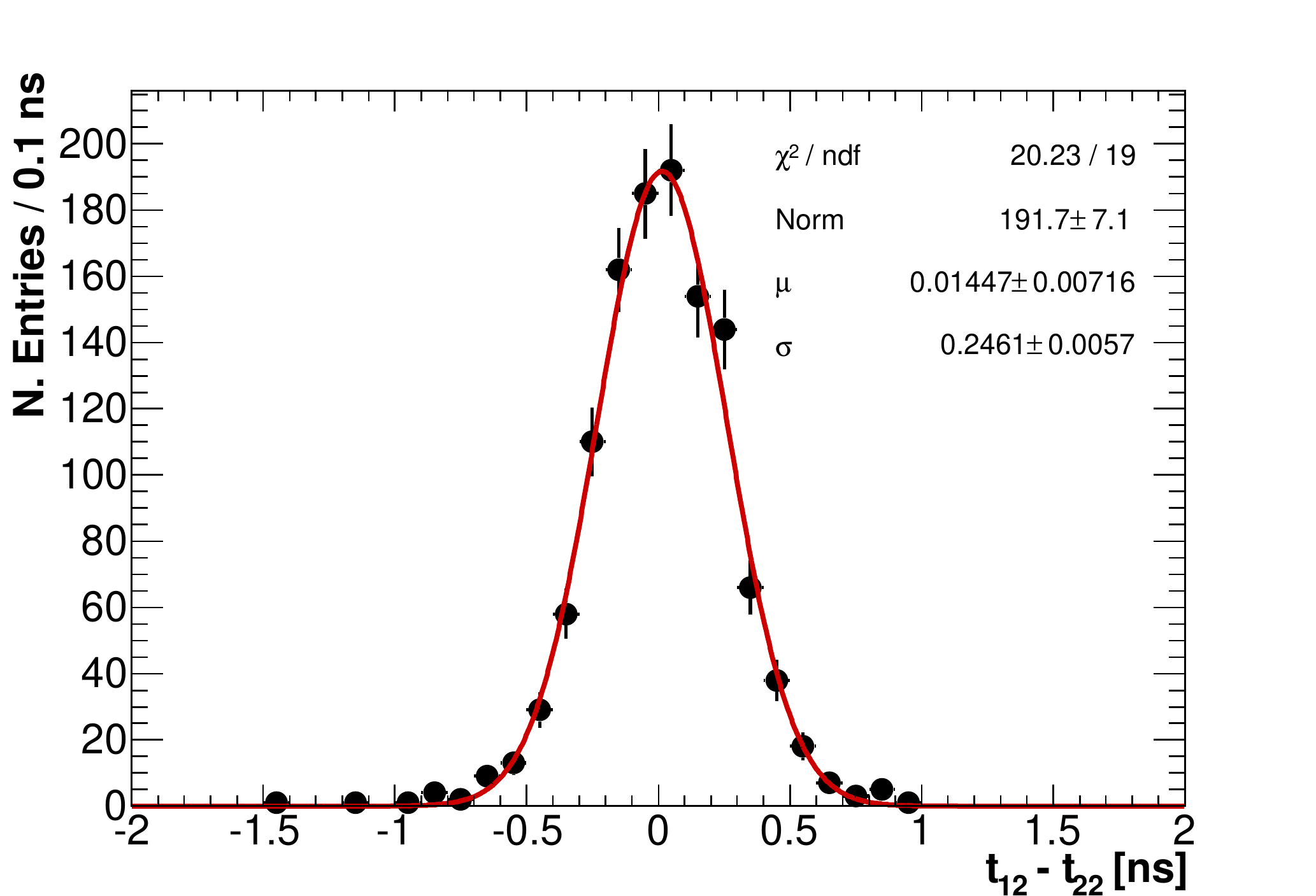}
\caption{Time residual distributions between signals in two adjacent crystals
    for $E_{beam}$ = 100 MeV (left) and $E_{beam}$ = 200 MeV (right).}
\label{fig:tdifference}
\end{center}
\end{figure}

Fig.~\ref{fig:slewinsitu} shows an example of the time-walk correction fit for $E_{beam}=100$ MeV.
Fig.~\ref{fig:tdifference} shows the $t_{12}-t_{22}$ distributions for beam energies of 100 and 200 MeV.

\subsection{Time resolution with MIPs and laser}
Measurements using minimum ionizing particles (MIPs) allow the time resolution at energies significantly below 100 MeV to be determined. For that, we used data collected with the cosmic trigger described in section \ref{sec:setup}. Selected events were required to have the energy deposition above 5 MeV in a column of 5 crystals, surrounded by two columns with energy depositions below 5 MeV in each. 


The time resolution has then been measured for a single crystal, using as $t_{start}$ the time of another crystal of the same column. The time resolution averaged over multiple tested pairs of crystals was ($312\pm9$)~ps for a mean energy deposition of ($23.1\pm0.3$)~MeV, estimated using the \texttt{GEANT4}-based simulation. 

The same procedure has been applied for the energy-weighted time average of two crystals, using as $t_{start}$ the time average of the two other crystals in the same column. The time resolution corresponding to the mean energy deposition of ($46.2\pm0.6$)~MeV was  ($262\pm7$)~ps. 

In order to estimate the contribution to the timing uncertainty from the photosensor, FEE, and digitizer, the data collected with the laser trigger at the APD gain $G=75$ have been used. 

Ten out of 25 crystals have been chosen to provide a reference time
\begin{equation}
t_{start} ~=~ \frac{\sum\limits_{j=1}^{10}t_jE_j}{\sum\limits_{j=1}^{10}E_j}.
\end{equation}

\begin{figure}[htbp]
\begin{center}
\includegraphics[scale=0.6]{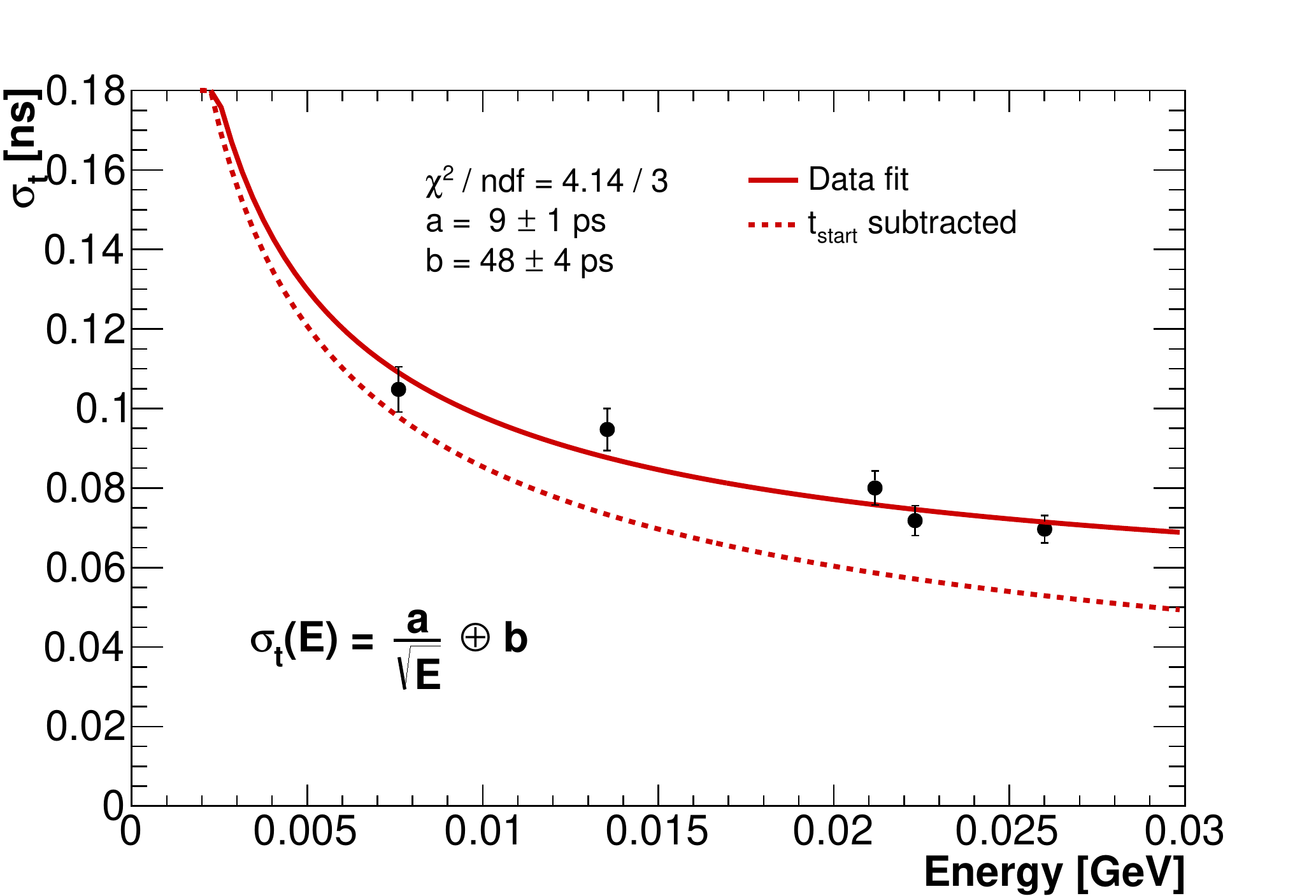}

\caption{Time resolution for laser signals as a function of the equivalent deposited energy.}
\label{fig:laser}
\end{center}
\end{figure}

The rest fifteen crystals have been split into 5 groups of 1, 2, 3, 4, and 5 crystals correspondingly. For each group, an energy-weighted mean time has been calculated as 
\begin{equation}
t_{n} = \frac{\sum\limits_{i=1}^n t_iE_i}{\sum\limits_{i=1}^n E_i} 
\end{equation}
and the widths of the distributions in $t_n - t_{start}$ have been plotted versus the mean total energy corresponding to the integrated charge in the crystals included in the group, as shown in Fig.~\ref{fig:laser}. 
%
%
A fit with the function:
\begin{equation}
\sigma_t(E) = \frac{a}{\sqrt{E/\mathrm{GeV}}} \oplus b,
\end{equation}
returns $a=(9\pm1)$ ps and $b=(48\pm4)$ ps.

\section{Conclusion}
\begin{figure}[htbp]
\begin{center}
\includegraphics[scale=0.6]{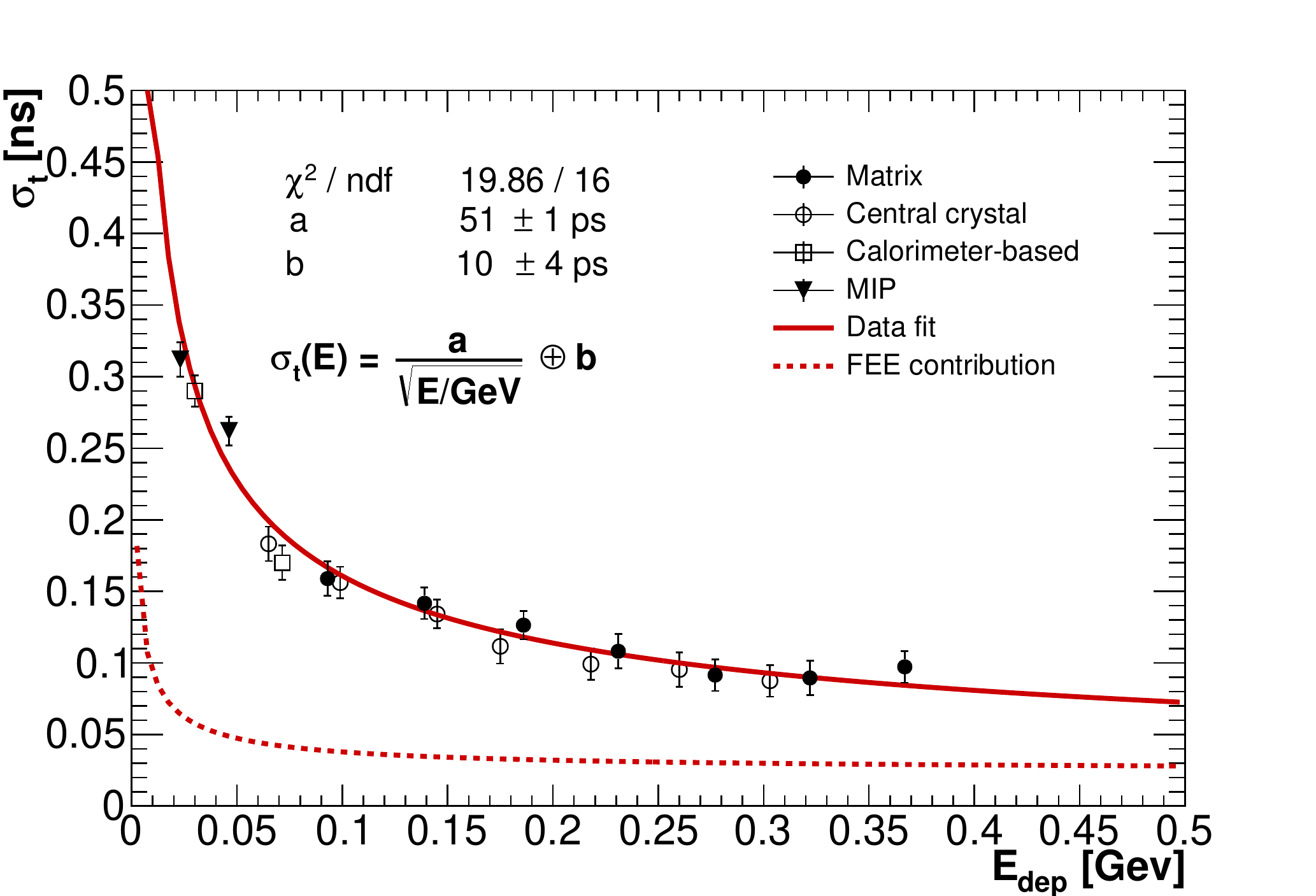}

\caption{Time resolution as a function of the deposited energy,
  obtained with scintillation counters (both for the central crystal and
  the entire matrix), with calorimeter-based technique and for MIP events. 
  The dashed line shows the FEE contribution obtained with the laser signal fit (Fig.~\ref{fig:laser}).}
\label{fig:reso_tot}
\end{center}
\end{figure}

The calorimeter time resolutions measured using different techniques are shown together in 
Fig.~\ref{fig:reso_tot}. Measurements with the external start are corrected for the timing jitter of the start signal. Measurements using the neighboring crystals method are converted to the single channel timing resolution assuming the resolution of both channels is the same.



The energy dependence of the timing resolution is parameterized as 
\begin{equation}
\label{eq_laser}
\sigma_t(E_{dep}) = \frac{a}{\sqrt{E_{dep} / \mathrm{GeV}}} \oplus b,
\end{equation}

with the stochastic term $a=(51\pm1)$ ps and the constant term $b=(10\pm4)$ ps 
determined from the fit. 

The time resolution of the LYSO calorimeter prototype at 100 MeV, $\sigma_t = (162 \pm 4)$~ps, amply satisfies the Mu2e calorimeter requirement $\sigma_t < 500$~ps~\cite{Mu2e_tdr}.

The time resolution also can be presented as the quadratic sum of three terms
\begin{equation}
\sigma_t = \frac{\tau_{s}}{\sqrt{{N_{p.e.}/\mathrm{MeV}}}} \oplus \sigma_{FEE} \oplus \sigma_{x},
\end{equation}
where the first term is due to the photo statistics and the emission time of the scintillator $\tau_s$, the second term corresponds to the timing jitter due to the photosensor and electronics, and the last term, $\sigma_{x}$, accounts for the shower length fluctuations, reconstruction and calibration-related terms.

For $\tau_{s}\rm{(LYSO)}=40 $ ns \cite{biblio:pdg} and the number of photoelectrons  estimated at N$_{\rm{p.e.}}$/MeV $=4100$,  
the first term at 100 MeV contributes around 63 ps. According to Eq.~\eqref{eq_laser}, at this energy, $\sigma_{FEE}$ = 38 ps.  
The remaining contribution, $\sigma_{x}\sim144$~ps, is dominated by the waveform reconstruction technique used.

Due to the limited energy coverage of our data is difficult to
  estrapolate this results to higher energy range ($>$ GeV) where the
  stochastic part of this dependence will be practically negligible.
  In order to reach timing close to the constant term i.e. O(10) ps,
  there will be need to improve both the front end electronics, that should
  demonstrate to have a rise time independent from the pulse height, and
  the fitting technique of the pulse shape.

\section*{Acknowledgments}
The authors are grateful to many people for the successful realization of the calorimeter prototype. In particular, we thank all the LNF mechanical shop for the realization of the support and the APD boxes. We also thank the whole BTF staff  for providing the beam time and helping us in getting a smooth running period. We express our warmest thanks to Luca Foggetta, for adjusting and tuning the beam to our detector needs.

\section*{References}

\bibliography{bibfile}

\end{document}